\documentclass[twocolumn]{emulateapj}
\usepackage{graphicx}
\usepackage{float}
\usepackage{mathrsfs}	
\usepackage{amsmath}
\usepackage{color}
\usepackage{hyperref}
\usepackage{multirow}
\usepackage{subfigure}

\begin{document}
\title{The discrepancy between Einstein mass and dynamical mass for SIS and power-law mass models}
\author{\mbox{Rui Li\altaffilmark{1,2,3,4}}}
\author{\mbox{Jiancheng Wang\altaffilmark{1,2,3,4}}}
\author{\mbox{Yiping Shu\altaffilmark{5,6}}}
\author{\mbox{Zhaoyi Xu\altaffilmark{1,2,3,4}}}

\altaffiltext{1} {Yunnan Observatories, Chinese Academy of Sciences, 396 Yangfangwang, Guandu District, Kunming, 650216, P. R. China,lirui@ynao.ac.cn, jcwang@ynao.ac.cn}
\altaffiltext{2} {University of Chinese Academy of Sciences, Beijing, 100049, P. R. China}
\altaffiltext{3} {Center for Astronomical Mega-Science, Chinese Academy of Sciences, 20A Datun Road, Chaoyang District, Beijing, 100012, P. R. China}
\altaffiltext{4} {Key Laboratory for the Structure and Evolution of Celestial Objects, Chinese Academy of Sciences, 396 Yangfangwang, Guandu District, Kunming, 650216, P. R. China}
\altaffiltext{5} {Purple Mountain Observatory, Chinese Academy of Sciences, 2 West Beijing Road, Nanjing, Jiangsu, 210008, China, yiping.shu@pmo.ac.cn}
\altaffiltext{6} {Institute of Astronomy, Madingley Road, Cambridge CB3 0HA, UK}

\begin{abstract}
We investigate the discrepancy between the two-dimensional projected lensing mass and the dynamical mass for an ensemble of 97 strong gravitational lensing systems discovered by the Sloan Lens ACS (SLACS) Survey, the BOSS Emission-Line Lens Survey (BELLS), and the BELLS for GALaxy-Ly$\alpha$ EmitteR sYstems (BELLS GALLERY) Survey. We fit the lensing data to obtain the Einstein mass, and use the velocity dispersion of the lensing galaxies provided by the Sloan Digital Sky Survey (SDSS) to get the  projected dynamical mass within the Einstein radius by assuming the power-law mass approximation. The discrepancy is found to be obvious and quantified by Bayesian analysis. For the singular isothermal sphere (SIS) mass model,  we obtain that the Einstein mass is 20.7\% more than the dynamical mass, and the discrepancy increases with the redshift of the lensing galaxies. For more general power-law mass model, the discrepancy still exists within $1\sigma$ credible region. The main reason of the the discrepancy could be mass contamination, including all invisible masses along  the line of sight. In addition, the measurement errors and the approximation of the mass models could lead to part of the discrepancy.
\end{abstract}


\section{Introduction}
Measuring the mass of galaxies is a fundamental work in galaxy astronomy. We know that a galaxy contains two main components: the stars and the dark matter. According to the $\Lambda$ cold dark matter theoretical model, the stellar component of a galaxy is embedded in a dark matter halo. In order to know the properties of the stellar component, we can use Stellar Population Synthesis (SPS) techniques to infer them. But as for the dark matter component, it emits no light or any other kinds of radios, so we have only a few ways to probe it. Among these ways, gravitational lenses must be the most powerful technique. It can constrain the total mass of the galaxies inside the Einstein radius  without knowing the the dynamical state and the nature of the matter. Another way to obtain the total mass of a galaxy is dynamical methods, for example, we can use the stellar dynamics to calculate the dynamical mass of the early-type galaxies (ETGs). After knowing the total mass and the stellar mass of a galaxy, we can get the mass of the dark matter component.

However, if we want to study some properties of the ETGs in detail, such as the inner mass profiles, the scaling relationships and the formation or the evolution of the galaxies, the mass estimations based on the gravitational lenses and the dynamics face some problems. The gravitational lenses can precisely constrain the total projected masses within the Einstein radius, but it can't provide us the detail distribution of the mass density profiles \citep{1985ApJ...289L...1F}. For the dynamical method, there is a degeneracy between the mass profiles and the stellar velocity dispersion tensors, known as the mass-anisotropy degeneracy (e.g., \cite{1993MNRAS.265..213G}), and it is also limited by the paucity of the bright kinematic traces in the outer regions of the ETGs.

A good method to break the degeneracy mentioned above is a joint analysis of gravitational lenses and stellar dynamics \citep{Treu02, 2004ApJ...611..739T, 2003ApJ...583..606K, 2006ApJ...649..599K}. This method use gravitational lenses to obtain the projected Einstein mass within the Einstein radius, then the Einstein mass is used to constrain the total mass of the galaxies by assuming the power-law mass model. Finally we can solve the spherical Jeans equation to determine the average mass-density slope. Over the past decade, several projects, such as the Sloan Lens ACS (SLACS) \citep{2008ApJ...682..964B}, the BOSS Emission-Line Lens Survey (BELLS) \citep{2012ApJ...744...41B}, the BELLS for the GALaxy-Ly$\alpha$ (BELLS GALLERY) \citep{2016ApJ...833..264S} and the Strong Lensing Legacy Survey (SL2S) (e.g., \cite{2012ApJ...761..170G}), have found more than 100 galaxy-scale strong gravitational lensing systems. Using these lensing systems and the joint analysis of gravitational lenses and stellar dynamics, \cite{2010ApJ...724..511A} figured out the mass-density slope of 73 SLACS samples and found that denser galaxies have steeper mass-density profiles. By combining the SLACS and BELLS samples, \citet{2012ApJ...757...82B} pointed out that the slope of the total mass-density profile has a linear relation with redshift, e.g., $\gamma=2.11-0.60z$. In the work of \cite{2014ApJ...786...89S}, based on the lensing measurements about the evolution of the mass-density profiles of ETGs, they used the SL2S samples to carry out a new test of the dry-merger scenario and suggested that the outer regions of massive ETGs grow through the accretion of stars and dark matter, while small amounts of dissipation and nuclear star formation conspire to keep the mass-density profile constant and approximately isothermal. The combination of gravitational lenses and stellar dynamics has also been used to study the cosmology. For example, assuming that the dynamical mass within the Einstein radius equals to the Einstein mass, \cite{2015ApJ...806..185C} constrained the equation of state to be $\omega=-1.15_{-1.20}^{+0.56}$.

However, in the joint analysis, the gravitational lenses and stellar dynamics are treated as two independent problems, and different potentials are used for the lens galaxies. It is shown that the joint analysis is not self-consistent. In order to overcome this shortcoming, \cite{2007ApJ...666..726B} developed a unifying framework for self-consistent analyses of ETGs. For any given galaxy potential, if the gravitational lenses data (i.e., the surface brightness distribution of the lensed images) and the stellar dynamics data (i.g., the surface brightness distribution and the line of sight projected moments of the lens galaxy) are provided, their method can find the best potential model using the Bayesian method. They also developed the CAULDRON algorithm to study SLACS samples \citep{2009MNRAS.399...21B,2011MNRAS.415.2215B}. The CAULDRON algorithm can also be used to study the late-type spiral galaxies \citep{2012MNRAS.422.3574B,2012MNRAS.423.1073B,2013MNRAS.428.3183D,2014MNRAS.437.1950B}.

Does the projected dynamical mass inside the Einstein radius indeed equal to the Einstein mass? As for the gravitational lensing images, their light could be affected by invisible mass contamination,  which could produce errors on the gravitational lens fitting results. Comparing the Einstein mass and dynamical mass for 27 SLACS samples, \cite{2007arXiv0706.3098G} found that the Einstein mass is about 6\% larger than the projected dynamical mass inside the Einstein radius for the SIS mass model, and they attributed this discrepancy to the line of sight mass contamination. Recently, \cite{2017arXiv170103418H} put forward that the dark energy inside the Einstein radius could partly offset the gravitational effect of both visible matter and dark matter, leading to an underestimation of the Einstein mass. In this work, we figure out the Einstein masses and the projected dynamical masses of 97 strong gravitational lensing systems and study the discrepancy between them. Throughout this paper, we adopt a fiducial cosmological model with $\rm \Omega_m = 0.274$, $\rm \Omega_{\Lambda} = 0.726$, and $H_0 \rm = 70\,km\,s^{-1}\,Mpc^{-1}$.

\section{date sets}
Our samples is composed of 97 strong gravitational lensing systems, where 57 come from SLACS \cite{2008ApJ...682..964B}, 25 come from BELLS \cite{2012ApJ...744...41B} and 15 come from BELSS GALLERY \cite{2016ApJ...833..264S}. All of the parameters, including the Einstein radius, the velocity dispersions, the redshifts of lensing galaxies, and the redshifts of background sources, are from their works too. We abandon two lensing systems of BELLS GALLERY because both of them have two lensing galaxies and we can't obtain their velocity dispersions. Notice that all of the three works mentioned above used the singular isothermal ellipsoid (SIE) mass model to fit the masses of the lensing galaxies, but here we use the SIS model and the power law model to study the mass discrepancy. However, we point out that this difference has no influence on our result, because in the lensing systems, the lensing images just relate to the total mass inside the Einstein radius and have no relationship with the mass models. The velocity dispersion is the average stellar velocity dispersion inside an aperture which is 1.5'' for SLACS and 1'' for BELLS and BELLS GALLERY.

\section{comparing Einstein mass and dynamical mass}
In this section,  we firstly introduce two methods of mass estimations, based on gravitational lenses and stellar dynamics, respectively. We then describe our Bayesian Method to quantify the discrepancy between the two mass estimations. Finally, we present our result of the parameters.

\subsection{methodology}
In a galaxy-galaxy gravitational lensing system, the light from the source will be deflected by the mass of lensing galaxy and generates arcs or multiple images. The locations of the observed images can help us to know the Einstein radius $\theta_E$.  The density inside the Einstein radius is the so called critical projected mass density,  which can be described as
\begin{equation}
\Sigma_{crit}=\frac{c^2}{4\pi G}\frac{D_s}{D_lD_{ls}}.
\end{equation}

The Einstein mass is $M_{ein}=\pi R_E^2 \Sigma_{crit}$, where $R_E=D_l \theta_E$ is the Einstein radius in the lens plane. Therefore, the $M_{ein}$ can be written as

\begin{equation}
M_{ein}=\frac{c^2}{4G}\frac{D_sD_l}{D_{ls}}\theta_E^2,
\end{equation}
where $D_l$ and $D_s$ are the angular-diameter distances of the lens and source, respectively. $D_{ls}$ is the angular-diameter distance between the lens and source.

\begin{figure*}[htbp]
\centering
\includegraphics[width=0.40\textwidth]{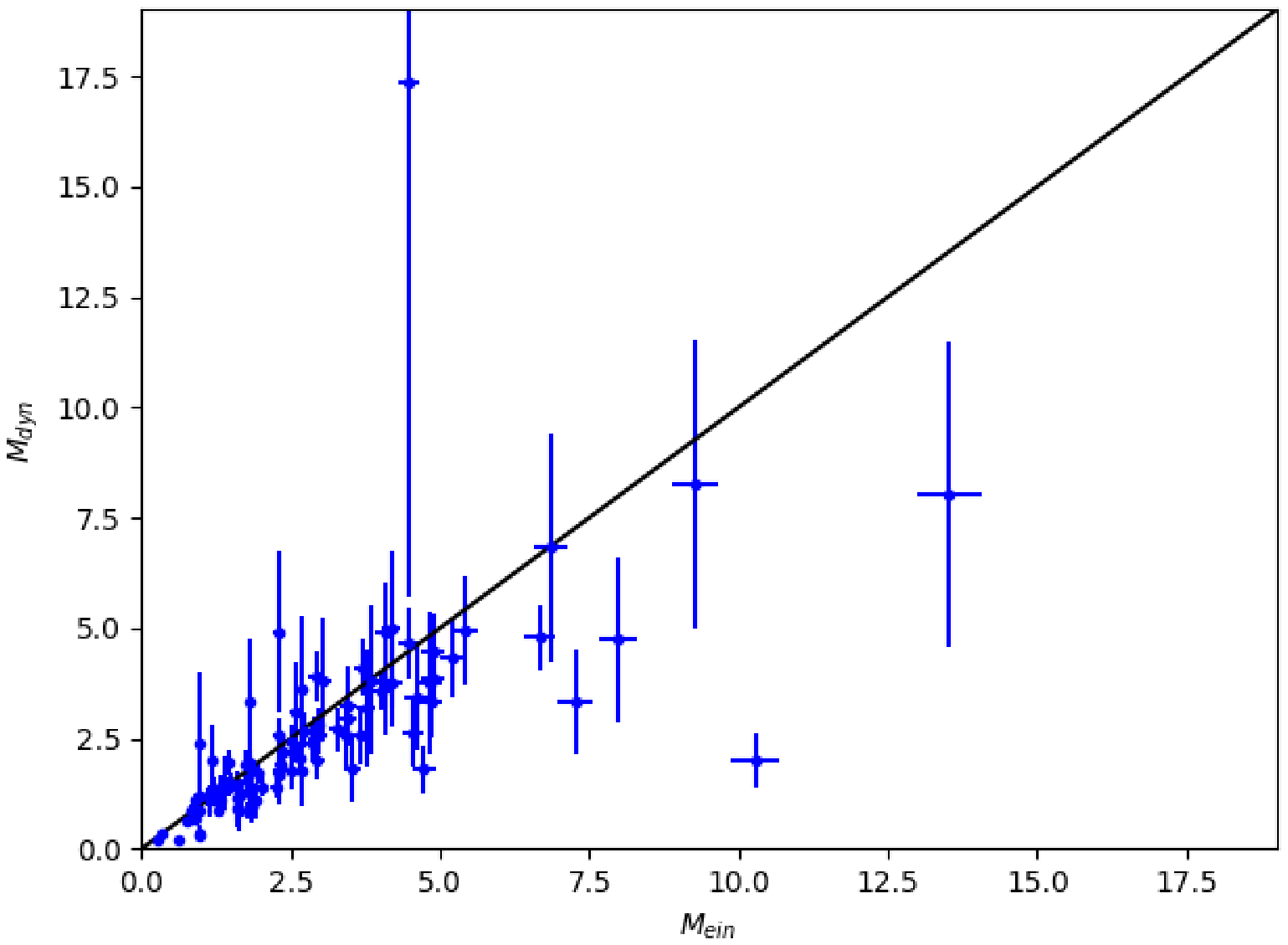}
\includegraphics[width=0.40\textwidth]{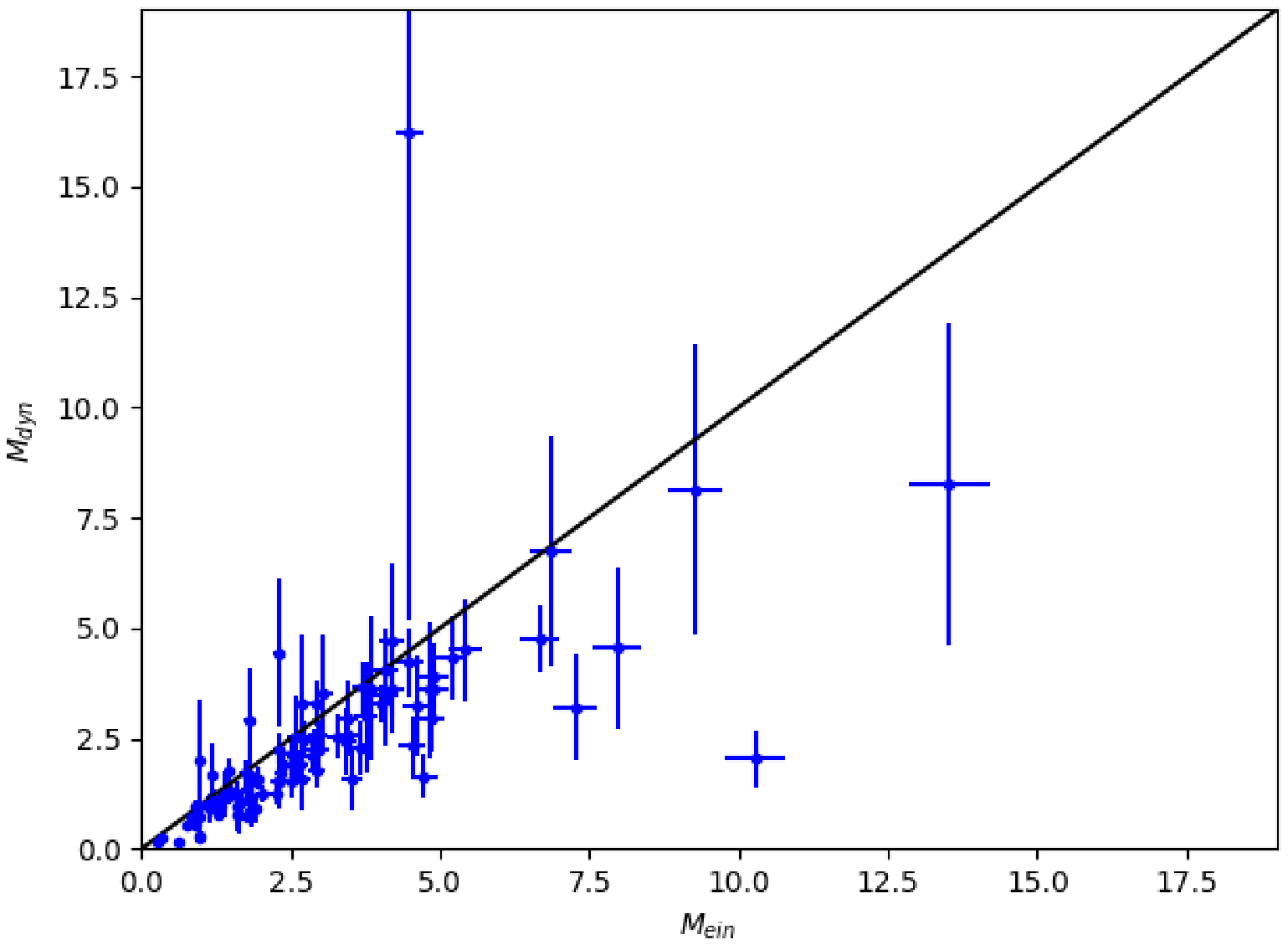}
\caption{\label{fig:illustration}
Comparison of the Einstein mass and the projected dynamical mass for for all the 97 lensing galaxies. The left figure is for the SIS mass model and the right figure is for the best fitting value of slope $\gamma$ of the power-law mass model. The lateral axis and the vertical axis are the Einstein mass and dynamical mass, respectively, they are both in units of $10^{11}M_\odot$.
}
\end{figure*}

SDSS provides us the velocity dispersion $\sigma_{ap}$ inside an aperture $\theta_{ap}$, where the aperture is 1.5'' for SLACS and 1'' for BELLS and BELLS GALLERY. We can use this to calculate the projected dynamical mass inside a circle. Assuming the power-law mass model $\rho\propto r^{-\gamma}$ for the lensing galaxy, we can write the dynamical mass  within the Einstein radius as \citep{Koopmans05, 2007arXiv0706.3098G, 2015ApJ...806..185C}

\begin{equation}
M_{dyn}(<\theta_E)=\frac{\pi}{G}\sigma_{ap}^2D_l\theta_E(\frac{\theta_E}{\theta_{ap}})^{2-\gamma}f(\gamma),
\end{equation}

where
\begin{equation}
\begin{split}
f(\gamma) &=-\frac{1}{\sqrt{\pi}}\frac{(5-2\gamma)(1-\gamma)}{3-\gamma}\frac{\Gamma(\gamma-1)}{\Gamma(\gamma-3/2)} \\
& ~~~\times (\frac{\Gamma(\gamma/2-1/2)}{\gamma/2})^2.
\end{split}
\end{equation}

For the SIS mass model, $\gamma=2$ and $f(\gamma)=1$, the projected dynamical mass within the Einstein radius has a simplified form

\begin{equation}
\begin{split}
M_{dyn}(<\theta_E) &=\frac{\pi}{G}\sigma_{ap}^2D_l\theta_E  \\
                   & =\frac{\pi}{G}\sigma_{ap}^2R_E.
\end{split}
\end{equation}

 Figure 1 shows the comparison of the Einstein mass and the projected dynamical mass for 97 lensing galaxies. The left figure is for the SIS mass model and the right figure is for the power-law mass model under the best fitting $\gamma$ (we will explain it in Subsection 3.2). The Einstein mass is calculated from the Einstein radius, while the dynamical mass is calculated from the velocity dispersion inside an aperture. We don't have the errors of the Einstein radius, but \cite{2008ApJ...682..964B} estimated the systematic uncertainty on Einstein radius measurements to be about 2\%. Therefore using this uncertainty, we find that the errors of the Einstein mass is about 5\%. The straight line in Figure 1 represents the dynamical mass equaling to the Einstein mass. We find that most of the points locate under the straight line. Therefore, the Einstein mass is a little larger than the dynamical mass. We also find that the galaxies with larger masses have larger mass discrepancy.

Next, we use the Bayesian method  described in \cite{2010arXiv1008.4686H} to quantify the discrepancy between the two mass estimations and find the confidence regions of the parameters. Within the Einstein radius, we suppose
\begin{equation}
M_{ein}=(1+a)M_{dyn},
\end{equation}
where $a$ is a free parameter in our Bayesian analysis. For the SIS mass model, if we know the Einstein radius, we can get the $\sigma_{ap}$ by
\begin{equation}
\sigma_{ap}=\sqrt{\frac{\Sigma_{crit}GR_E}{1+a}}.
\end{equation}
Here, we denote the velocity dispersion calculated from the Einstein radius using equation(7) as $\sigma_{pre}$, the velocity dispersion calculated from SDSS spectroscopic data as $\sigma_{obs}$. Therefore, the likelihood function is
\begin{equation}
ln{\cal L} =-\frac{1}{2}\sum_{i=1}^{i=97}[\frac{(\sigma_{pre}^i-\sigma_{obs}^i)^2}{s_i^2}+ln(2\pi s_i^2)],
\end{equation}
where
\begin{equation}
s_i^2=\delta^2+(\sigma_{err}^i)^2,
\end{equation} 
$\delta$ represents the intrinsic scatter between $\sigma_{pre}$ and $\sigma_{obs}$, and we treat it as a free parameter in our Bayesian analysis. $\sigma_{err}^i$ is the measurement errors on $\sigma_{obs}^i$. Now, the posterior probability is
\begin{equation}
p(a,\delta | {\left\lbrace  \theta_E^i \right\rbrace })={\cal L}({\left\lbrace  \theta_E^i \right\rbrace } | a,\delta)p(a,\delta)
\end{equation}
We choose a flat prior $p(a,\delta)$ for $a$ and $\delta$ ($-1<a<1$ and $0<\delta<50$). With  these two free parameters, we use a Python implementation named Emcee \citep{2013PASP..125..306F} to do our Bayesian analysis.

\subsection{result}

Figure 2 shows the posterior probability distributions of $a$ and $\delta$ for the SIS mass model.  The results are $a=0.207_{-0.046}^{+0.041}$, $\delta=30.3_{-4.0}^{+3.7}$, indicating that there is indeed a discrepancy between the Einstein mass and the projected dynamical mass.

\begin{figure}[H]
\centering
\includegraphics[width=0.45\textwidth, height=0.35\textwidth]{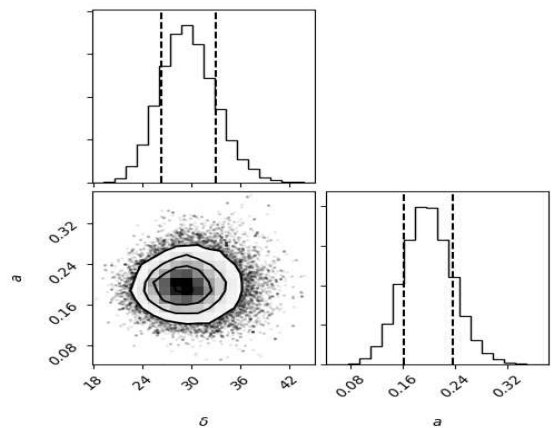}
\caption{\label{fig:illustration}
Assuming a SIS mass model, we give the posterior probability distributions of $\delta$ and $a$. The regions between the two dashed lines are the $1\sigma$ credible regions.
}
\end{figure}

Now we let that $a$ changes with the Einstein mass or with the Einstein radius (in kpc) by
\begin{eqnarray}
a(M_{ein})=a_0+a_{M}\left(\frac{M_{ein}}{10^{11}M_{\sun}}-3.0\right),\\
a(R_{ein})=a_0+a_{R}(R_{ein}-5.17),
\end{eqnarray}
where $M=3*10^{11}M_\odot$ and $R_{ein}=5.17kpc$ which, are the mean values of the Einstein mass and Einstein radius for 97 lenses, respectively.  After Bayesian analysis, we get $a_0=0.207_{-0.035}^{+0.036},\ a_M=0.067_{-0.019}^{+0.020}$ for equation(12), and $a_0=0.219_{-0.036}^{+0.037},\ a_R=0.065_{-0.017}^{+0.018}$ for equation(13). We can see that the discrepancy between the Einstein mass and dynamical mass still exists. The evolution factor $a_M$ and $a_R$ are very small, but it is noted that the Einstein masses of our sample range from $0.26\times10^{11}M_\odot$ to $13.52\times10^{11}M_\odot$, and the Einstein radius range from $1.14kpc$ to $14.8kpc$.  Therefore, the parameter $a$ of the biggest galaxy is at least $0.3$ or more larger than that of the smallest galaxy, we state that $a$ evolves with the Einstein mass.

Next, we let that $a$ changes with the redshifts of lensing galaxies and background sources as
\begin{eqnarray}
a(z_{lg})=a_0+a_{lg}(z_{lg}-0.3),\\
a(z_{bg})=a_0+a_{bg}(z_{bg}-1.07),
\end{eqnarray}
where $z_{lg}$ is the lensing redshift, and $z_{bg}$ is the source redshift. $z_{lg}=0.30$ and $z_{bg}=1.07$ are the mean redshifts of the lensing galaxies and background sources, respectively. We find that the results are $a_0=0.216_{-0.034}^{+0.037}, \ a_{lg}=0.869_{-0.255}^{+0.260}$ for equation(14), and $a_0=0.235_{-0.043}^{+0.048}, \ a_{bg}=-0.157_{-0.089}^{+0.095}$ for equation(15). Here, we still find the discrepancy between the Einstein mass and dynamical mass. Additionally, we find that $a$ strongly depends on lenses redshift, galaxies with higher redshift have bigger discrepancy between the Einstein mass and the projected dynamical mass.

It is found that the mean Einstein radius of BELLS (5.70kpc) and BELLS GALLERY (8.11kpc) are larger than that of SLACS (4.17kpc), while their redshifts are higher. Therefore, the relation between $a$ and $R_{ein}$ and the relation between $a$ and $z$ could affect each other and give us imprecise constrains for the parameters.   In order to examine this guess, we assume that
\begin{equation}
a(z_{lg}, R_{ein})=a_0+a_{R}(R_{ein}-5.17)+a_{lg}(z_{lg}-0.3).
\end{equation}
The Bayesian results are $a_0=0.224_{-0.035}^{+0.037}$, $a_R=0.047_{-0.022}^{+0.022}$, $a_{lg}=0.371_{-0.305}^{+0.320}$, implying that $a$ still increases with the Einstein radius and the lenses redshift within $1\sigma$ error.

\begin{table*}
\caption{\label{tab:result} Summary of the constraints on the
parameters for SIS mass model and power-law mass model (see text for definitions).}
\begin{center}%{\scriptsize%\footnotesize
\begin{tabular}{lllllllll}\hline\hline
Model(equation)  & parameters \{$\delta$, $a$, $a_0$, $a_M$, $a_R$, $a_{lg}$, $a_{bg}$, $\gamma$, $\gamma_0$, $\gamma_z$\}  \\
\hline
SIS   & $\delta= 30.3^{+3.7}_{-4.0}$, \ $a= 0.207^{+0.041}_{-0.046}$ \\
SIS (equation 11) & $\delta= 26.1^{+3.5}_{-3.0}$, \ $a_0= 0.207^{+0.036}_{-0.035}$, \ $a_M= 0.067^{+0.020}_{-0.019}$ \\
SIS (equation 12) & $\delta= 25.6^{+3.4}_{-3.1}$, \ $a_0= 0.219^{+0.037}_{-0.036}$, \ $a_R= 0.065^{+0.018}_{-0.017}$ \\
SIS (equation 13) & $\delta= 25.6^{+3.5}_{-3.1}$, \ $a_0= 0.216^{+0.037}_{-0.034}$, \ $a_{lg}= 0.869^{+0.255}_{-0.260}$ \\
SIS (equation 14) & $\delta= 28.0^{+3.5}_{-3.3}$, \ $a_0= 0.235^{+0.048}_{-0.043}$, \ $a_{bg}= 0.157^{+0.095}_{-0.089}$ \\
SIS (equation 15) & $\delta= 25.2^{+3.3}_{-3.1}$, \ $a_0= 0.224^{+0.037}_{-0.035}$, \ $a_R= 0.047^{+0.022}_{-0.022}$, \ $a_{lg}= 0.371^{+0.320}_{-0.305}$ \\
\hline

Power-law   & $\delta= 28.2^{+3.8}_{-4.1}$, \ $a= 0.696^{+0.474}_{-0.352}$, \ $\gamma= 1.751^{+0.101}_{-0.095}$ \\
  %\hline
Power-law (equation 16 and 17)  & $\delta= 26.3^{+3.7}_{-4.0}$, \ $a_0= 0.437^{+0.212}_{-0.170}$, \ $a_{lg}= 00.881^{+0.564}_{-0.798}$, \ $\gamma_0= 1.837^{+0.106}_{-0.095}$, \ $\gamma_z= 0.144^{+0.269}_{-0.376}$ \\
  %\hline

\hline\hline
\end{tabular}
\end{center}
\end{table*}

Apart from the SIS mass model, we extend our research to the more general power-law mass model. We treat $a$ and the average mass-density slope $\gamma$ as free parameters,  then we get the results of $a=0.696_{-0.352}^{+0.474}$ and $\gamma=1.751_{-0.095}^{+0.101}$. Here, the discrepancy between the Einstein mass and the projected dynamical mass is still found within $1\sigma$ credible region, and it is even bigger than that of SIS mass model.  However, as shown in Figure 4, we find a degeneracy between parameter $a$ and the average mass-density slope $\gamma$. It is also noted that our best-fitting $\gamma$ is  smaller than that of previous works, we think it is the degeneracy between $a$ and $\gamma$ that leads to the smaller $\gamma$ value. 

 \cite{2012ApJ...757...82B} pointed out that the average slope $\gamma$ evolves with redshift. Here we also find that $a$ increases with the lenses redshift, so the degeneracy may be caused by these two evolutions. Now we let that $a$ and $\gamma$ change with redshift at the same time:

\begin{eqnarray}
a(z_{lg})=a_0+a_{lg}(z_{lg}-0.3),\\
\gamma=\gamma_{0}+\gamma_{z}(z-0.3).
\end{eqnarray}
The results are $a_0= 0.437^{+0.212}_{-0.170}$, $a_{lg}= 00.881^{+0.564}_{-0.798}$, $\gamma_0= 1.837^{+0.106}_{-0.095}$, and $\gamma_z=  0.144^{+0.269}_{-0.376}$. The posterior probability distributions of all parameters are shown in Figure 5. We find that the degeneracy between $a$ and $\gamma$ still exists. In our study, $\gamma_z$ is positive, which is inconsistent with previous works \citep{2012ApJ...757...82B, 2013ApJ...777...98S}. However, our result have a wide credible region, we do not reject a negative value for $\gamma_z$.

\begin{figure}[H]
\centering
\includegraphics[width=0.45\textwidth, height=0.35\textwidth]{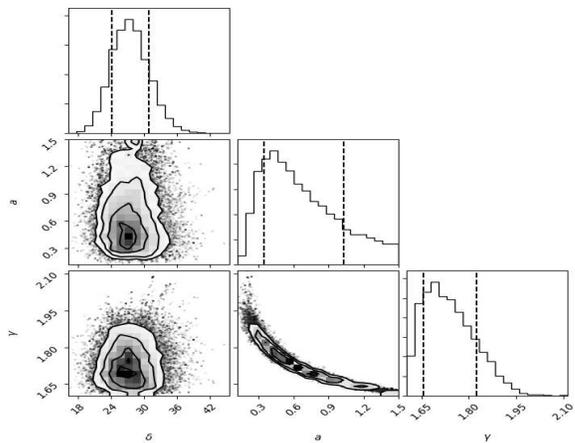}
\caption{\label{fig:illustration}
Assuming the power-law mass model, we give the posterior probability distributions for $\delta$, $a$ and $\gamma$. The regions between the two dashed lines are the $1\sigma$ credible regions.
}
\end{figure}

\begin{figure*}[htbp]
\centering
\includegraphics[width=0.9\textwidth, height=0.7\textwidth]{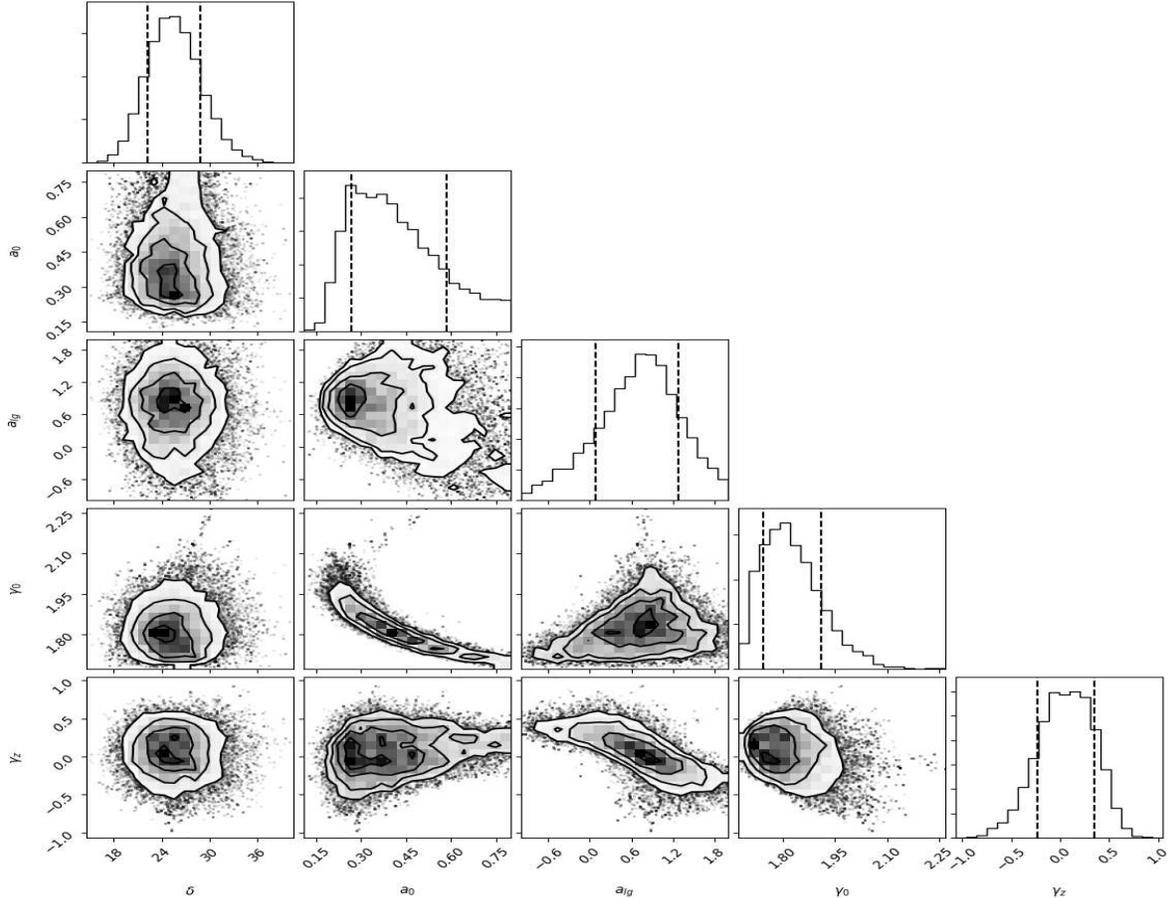}
\caption{\label{fig:illustration}
Assuming the power-law mass model and letting $a$ and $\gamma$ change with the redshift of lensing galaxies, we give the posterior probability distributions for $\delta$, $a_0$, $a_{lg}$, $\gamma_0$ and $\gamma_z$. The regions between the two dashed lines are the $1\sigma$ credible regions.
}
\end{figure*}

\section{discussion}
In this section, we will discuss the reasons of the discrepancy between Einstein mass and dynamical mass, including the invisible mass contamination and many other factors.

In the work of \cite{2007arXiv0706.3098G}, by comparing the Einstein mass and the projected dynamical mass of 27 gravitational samples of SLACS, they found that the lensing method overestimates the galaxy masses. Their result is $M_{ein}=1.06\pm0.08M_{dyn}$ for the SIS mass model. For other mass models, the difference was even larger. They attributed this difference to the line of sight mass contamination. Additionally, \cite{2006ApJ...641..169M}, \cite{2006ApJ...646...85W} and \cite{2007ApJ...660L..31M} also found the significant line of sight effects on strong gravitational lensing systems.  In this paper, we also state that the line of sight mass contamination is the main factor of the discrepancy between the Einstein mass and the projected dynamical mass. Additionally, we find that the discrepancy increases with the Einstein radius and the redshift of the lens galaxies. This tendency is easy to understand. If a lens galaxy has larger Einstein radius and higher redshift, the line of sight mass contamination is more likely to be larger and leads to larger discrepancy between the Einstein mass and the projected dynamical mass.

As mentioned above, the Einstein radius (or Einstein mass) has a certain effect on the discrepancy, which indicates that the discrepancy could be from the lens galaxies themselves or the surroundings of the lens galaxies. Here, we point out that a galaxy has its own best fitting value of slope $\gamma$, if we use the average $\gamma$ value for all the 97 galaxies, it could affect the the relation between Einstein mass and projected dynamical mass.  We also find that a systematic change of $\gamma$ would have a systematic effect on the discrepancy. We let $\gamma$ changes from $1.6$ to $2.4$ and use the Bayes method we mentioned above to calculate the value of $a$ for different $\gamma$. The relation between $a$ and $\gamma$ is shown in Figure 5. We try to use the quadratic function to fit the points and find the lowest point ($\gamma$,$a$) is ($2.010$, $0.111$). However, as we can see in Figure 5, the lowest point of the quadratic curve could underestimate the value of $a$. Additionally, based on the cold dark matter (CDM) model, some analytical calculations and numerical simulations  suggest that the substructures around the Milky Way are much more numerous than the observed numbers of satellite galaxies \citep{1999ApJ...522...82K, 2011MNRAS.415L..40B} and the same situation should appear in Early-type galaxies(ETGs) too. If this conception is right, we state that the invisible substructures of  ETGs within the Einstein radius would not lead to the discrepancy, because we use the average mass-density profiles in our study.

\begin{figure}[H]
\centering
\includegraphics[width=0.45\textwidth, height=0.35\textwidth]{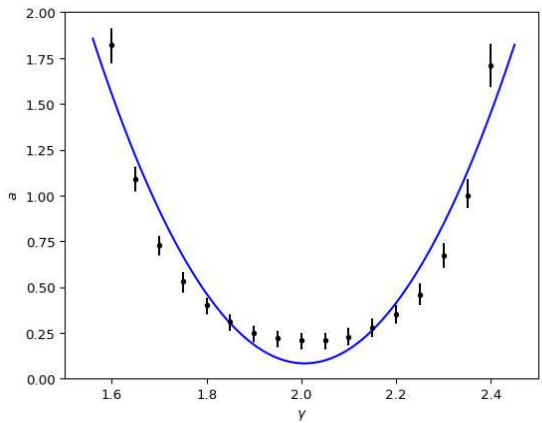}
\caption{\label{fig:illustration}
The discrepancy between Einstein mass and the projected dynamical mass for different $\gamma$ value. The blue line is the best fitting line.
}
\end{figure}

The dynamical mass estimated by the velocity dispersion is only an approximation. The uncertainty of this method is at the level of 20\%. It is noted that even the uncertainty is higher, the dots in Figure 1 should evenly distribute on both sides of the line. But most of the dots locate under the line, implying that the larger uncertainty of dynamical mass can't explain the discrepancy. We have not found that the dynamical mass estimated by velocity dispersion underestimates the mass of galaxy in other works.   Additionally, the SIS and power-law mass distribution we used in our study are elliptical and smooth mass density profile, we also neglect the anisotropy of the velocity distribution. This assumption may be too simple to describe the mass distribution of a real galaxy,  which may lead to part of the discrepancy. Therefore, if we want to get more accurate dynamical mass, more accurate mass model is needed. But in our study, at least for SIS mass model and power-law mass model, the discrepancy indeed exists. If we want to use these two kinds of mass models to make a joint analysis of gravitational lenses and dynamical information, it is worth to make a correction.  

 Some works have studied the influence of dark energy on local gravitational lensing systems. \cite{2011ICRC....5..223S} suggested that the cosmological constant has limited effects on lensing signals. \cite{2017arXiv170103418H}, pointed out that the dark energy acts as a concave lens, contrary to the convex lens of matter, and claimed that the influence of dark energy on galaxy-cluster lensing systems is significant, which could lead to the underestimation of the Einstein mass. In our result, the Einstein mass is about $20\%$ more than the dynamical mass for the SIS mass model, indicating that we may overestimate the Einstein mass. We think either the dark energy is too weak to cause the discrepancy on galaxy-scale or the theory in \cite{2017arXiv170103418H} does not square with observation.

\section{Conclusions}
In this work, we use 97 strong gravitational lensing systems selected from the SLACS, BELLS, and BELLS GALLERY, to quantify the discrepancy between the Einstein mass and the projected dynamical mass within the Einstein radius.  Using the SIS mass model, we find that the Einstein mass is 20.7\% more than the projected dynamical mass. The Einstein radius (or the Einstein mass) and the redshift of the lens galaxies have a certain effect on the discrepancy. When we use the power-law mass model to study the discrepancy, we find that the discrepancy still exists. We also find a degeneracy between $a$ and the average mass-density slope $\gamma$. The main factor of the discrepancy could be the line of sight mass contamination. We find that a change of $\gamma$ would have an effect on the discrepancy. The dark energy could offset partly gravitational influence for local lensing systems and cause an underestimation of the Einstein mass \citep{2017arXiv170103418H}. However, our result is not consistent with their conclusion, indicating that the dark energy is not the main factor to cause the discrepancy between the Einstein mass and the projected dynamical mass on galaxy scale.  Finally, when the SIS and power-law mass model are used in the joint analysis of gravitational lenses and dynamical information, we suggest that a correction to the Einstein mass is needed.

\section{Acknowledgements}
We acknowledge the financial support from the National Natural Science Foundation of China 11573060 and 11661161010. Y.S. has been supported by the National Natural Science Foundation of China 
(No. 11603032 and 11333008), the 973 program (No. 2015CB857003), and 
the Royal Society – K.C. Wong International Fellowship (NF170995).

\end{document}